# Development of a Causal Model for Improving Rural Seniors' Accessibility: Data Evidences

**Ke Li, Shizhe Li, and Ruwen Qin, Ph.D., A.M. ASCE**

Department of Civil Engineering, Stony Brook University, Stony Brook, NY 11794, email: ke.li.1@stonybrook.edu

**ABSTRACT**

Seniors residing in rural areas often encounter limited accessibility to opportunities, resources, and services. This paper introduces a model proposing that both aging and rural residency are factors contributing to the restricted accessibility faced by rural seniors. Leveraging data from the 2017 National Household Travel Survey, the study examines three hypotheses pertaining to this causal model. Multiple causal pathways emerge in the data analysis, with mobility identified as a mediator in one of them. The study further identifies specific challenges faced by rural seniors, such as the reduced accessibility in reaching medical services and assisting others. These challenges stem primarily from aging and geographic obstacles that not only diminish their willingness to travel but also restrict more in the group from choosing transportation modes with higher mobility. The insights gained from this study serve as a foundation for devising effective methods to enhance transportation accessibility for seniors in rural areas.

**INTRODUCTION**

People have unequal accessibility to spatially distributed opportunities, resources, and services that they intend to reach. Seniors in rural areas are a group with more restricted accessibility for multiple possible reasons. To promote equitable accessibility for this group, it is important to identify the root causes for their reduced accessibility. Compared to younger adults (aged 16 to 64) in rural areas, rural seniors have inherently different needs to access spatially distributed destinations. For example, a local clinic may no longer meet their special need for medical services. Due to the lower population density in rural areas, the cost for rural seniors to access certain destinations is more expensive than their counterparts in urban areas. That is, aging and rural residency are part of the reasons for the reduced accessibility of rural seniors. Aging and rural residency influence people's accessibility through other causal pathways too. For example, some rural seniors cease driving once it becomes impractical. Their heavy reliance on automobiles and the increased restriction for them to choose this transportation mode also reduce rural seniors' accessibility. That is, mobility is part of a causal pathway and it is a mediator of the causal relationship studied in this paper. Ignoring this causal pathway would underestimate the impact of aging and rural residency to the accessibility of rural seniors.

The study of this paper is motivated by the complexity of the causal relationship between the demographic characteristics of rural seniors and their accessibility. Specifically, it aims to determine how aging and rural residency negatively impact rural seniors' accessibility via multiple causal pathways and verify if mobility is a mediator. As such, improved mobility for rural seniors can effectively enhance the accessibility for this group. The National Household Travel Survey (NHTS) dataset (NHTS 2017) is used in this study, which is the only source of national data to study personal travel behavior. However, related work indicates that the selection of metrics for accessibility and mobility is affected by the study context and data availability. Therefore, discussions of the study in this paper are driven by the following two questions. What metrics of accessibility and mobility can be derived from the NHTS dataset? How can the assumed causal relationship be verified with evidences extracted from the NHTS dataset?

The remainder of the paper is organized as follows. The literature is reviewed to identify the foundation for this study and the gaps to be filled. Then, the research methodology is introduced, which is followed by the result discussion. In the end, the paper summarizes the major findings and recommendations for future work.

**THE LITERATURE**

**Accessibility and Mobility.** Accessibility is usually defined as the ability to reach the intended destinations (Litman 2003). According to Geurs and van Wee (2004), the four components contributing to accessibility encompass transportation, land use, individual factors, and time cost. Nonetheless, incorporating all these components as direct indicators of accessibility is challenging (Pyrialakou et al. 2016). Boisjoly and El-Geneidy (2017) reviewed various travel accessibility indicators such as the count of accesses to specific purposes within a defined time frame, which are influenced by factors like land use, opportunity distribution, and mobility. They also introduced two location-based measures of accessibility, namely the gravity-based measure and the cumulative opportunity measure. These measures have reached a level of maturity, extensively discussed in numerous research papers (El-Geneidy and Levinson 2006; Scott and Horner 2008; Casas 2007).

Mobility is defined as the movement of people and goods (Litman 2003). This concept underscores the act of movement rather than a mere means to a destination. Mobility was assessed using travel surveys to quantify person miles, ton-miles, and travel speeds. Furthermore, traffic data were utilized to gauge the average speeds of both automobiles and transit vehicles (Litman 2003). Mobility cannot be measured by a definitive or singular metric (Pyrialakou et al. 2016). Therefore, multiple indicators were developed, including travel mode distribution, travel frequency, and vehicle ownership (Pucher and Renne 2005; Jansuwan et al. 2013). To assess transportation disadvantages related to mobility, Kamruzzaman and Hine (2011) developed a comprehensive indicator called "participant index" (PI). It combines various elements, including the number of destinations visited, travel distance, space, travel frequency, types of activities, and duration. The mobility of individuals who do not drive can be enhanced by improved public transit infrastructure, reduced time required to reach transit stations, and increased proximity to their intended destinations (Case 2011).

**Disadvantage of Seniors in Rural Areas.** Martens (2015) introduced a framework for assessing accessibility offered by a transport-land use system and the potential mobility facilitated by the transport system. This study indicates that the improved accessibility is partially attributed to the enhanced mobility. Actually, mobility data were employed to assess accessibility (Mittal et al. 2023). However, the popularly used gravity-based measurements do not adequately account for the influence of mobility on accessibility. Both mobility and proximity play pivotal roles in enhancing accessibility, but they often exist in a trade-off relationship (Grengs et al. 2010). In areas where the origin and the destination are close (high proximity), travel speeds typically tend to be slower (low mobility). Therefore, accessibility measures should take into account the combined influence of location- and mobility-related factors.

Rural seniors are often perceived as a disadvantaged group, and their demographic characteristics are considered as reasons. For example, "young elderly" (aged 65-75) and the "old elderly" (aged 75+) have different travel patterns and expectations on accessibility (Alsnih and Hensher 2003). When aiming to enhance the accessibility of seniors, it is imperative to consider their mental and physical needs. Accessibility varies significantly among different groups of travelers, such as individuals with mobility impairments attributed to aging and those reliant on public transit (Márquez et al. 2019). In the United States, automobiles are the predominant travel mode in rural areas (Pucher and Renne 2005), and seniors particularly rely on automobiles for travel (Kim 2011). Rural seniors' mobility for daily travel needs is restricted when it comes to public transit and walking (Ravensbergen et al. 2021). Rural areas are typically less developed than urban areas from the view of accessibility to opportunities, resources, and services (Vitale Brovarone and Cotella 2020).

**METHODOLOGY**

**Hypotheses About the Causal Relationship.** This paper posits a causal relationship wherein the attributes of rural seniors are factors (i.e., independent variables) causing reduced accessibility (i.e., the dependent variable). While the interested relationship has multiple causal pathways, this paper examines two, as shown in Figure 1. The first causal pathway goes directly from the group's attributes to its accessibility. In the second pathway, the attributes impact accessibility via mobility, a mediator variable. Three hypotheses delineated below are underlying this causal relationship:

H1: Rural seniors' accessibility is directly impacted by their inherently altered needs for accessing opportunities, resources, and services, as well as their rural residency.

H2: Reduced mobility among rural seniors stems from their constrained choices of travel modes and decreased travel willingness.

H3: The reduced mobility among rural seniors is another reason that rural seniors experience reduced accessibility.

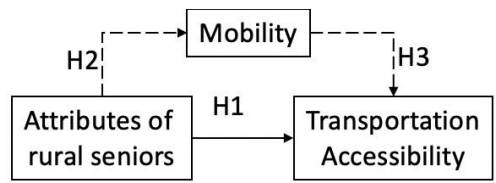

**Figure 1. The proposed causal model of accessibility for rural seniors.**

**The NHTS Dataset.** This study examines the hypothesized causal relationship in Figure 1 by extracting evidences from the 2017 National Household Travel Survey (NHTS 2017), which contains a completed survey from 129,696 households and 923,572 person trips. This study defines its scope of data analysis by concentrating on:
- Four primary travel modes: automobiles, bicycles, walking, and public transit
- Six key travel purposes: home, work, medical service, shopping, recreational activities, and transporting someone.
- Local travel with distance being within 75 miles (Pucher and Renne 2005).
- Two age groups: travelers from 16 to 64 years are defined as younger adults (Yadults), and those aged 65 or older are defined as seniors.
- Two regions of household residency: urban and rural based on home address and 2014 TIGER/Line Shapefile (NHTS 2017).

**Attributes and Measures.** Attributes or measures are defined for examining the influences among variables in the causal model. Age and residency area are selected as the demographic attributes for characterizing travelers. Consequently, two additional reference groups, rural Yadults, and urban seniors, are defined. Comparing rural seniors to rural Yadults facilitates the measurement of aging-induced changes while comparing them to urban seniors allows for the measurement of location-induced changes. This study chooses travel distance and travel time to intended destinations as measures of accessibility. Because both are random variables, their 75th percentiles conditioned on a specific travel purpose are used as indicators of the ease to reach an intended destination. This study chooses four indicators as the basis for evaluating mobility, including the distribution of trips by travel modes, travel frequency, travel speed, and time to access public transit stations. Notably, travel frequency, often expressed as the number of trips per person per day, is the principal indicator for measuring mobility in numerous prior studies (Pucher and Renne 2005; Szeto et al. 2017). Although factors like congestion and travel miles per person also hold significance for assessment (Litman 2003), this study selects travel speed and time to access public transit considering the data coverage of NHTS.

## DATA ANALYSES

**The Direct Impact of Rural Residency on Accessibility.** Hypothesis H1 asserts that the increased travel distance for rural seniors to reach their intended destinations, in contrast to their urban counterparts, is a contributing factor to their reduced accessibility. To verify this hypothesis, this study compares rural seniors to urban seniors on their cumulative distribution functions (CDFs) of travel distance, $F_{D,RSr}(x)$ and $F_{D,USr}(x)$, for each specific travel purpose. These two groups are also compared in terms of their travel time CDFs, $F_{T,RSr}(x)$ and $F_{T,USr}(x)$. Statistics of travel time and travel distance are further summarized in Table 1.

**Table 1. Statistics of travel distance and travel time by groups and travel purposes.**

| Purpose | $D_{0.5}$ (mile) | | | $D_{0.75}$ (mile) | | | $T_{0.5}$ (min) | | | $T_{0.75}$ (min) | | |
|---|---|---|---|---|---|---|---|---|---|---|---|---|
| | RSr | RYa | USr | RSr | RYa | USr | RSr | RYa | USr | RSr | RYa | USr |
| Home | 7.2 | 8.2 | 2.7 | 14.1 | 16.0 | 6.2 | 15.0 | 16.0 | 15.0 | 30.0 | 30.0 | 25.0 |
| Work | 7.7 | 9.4 | 4.4 | 16.0 | 19.4 | 10.3 | 15.0 | 18.0 | 15.0 | 30.0 | 30.0 | 30.0 |
| Medical | 13.3 | 12.1 | 5.0 | 22.9 | 20.4 | 11.2 | 30.0 | 23.0 | 20.0 | 45.0 | 32.0 | 30.0 |
| Shopping | 4.3 | 4.1 | 2.3 | 10.6 | 10.4 | 4.8 | 12.0 | 10.0 | 10.0 | 20.0 | 20.0 | 17.0 |
| Recreational | 4.7 | 5.7 | 3.1 | 11.2 | 13.8 | 7.9 | 15.0 | 15.0 | 15.0 | 17.0 | 30.0 | 25.0 |
| Transport someone | 7.0 | 6.0 | 3.6 | 14.9 | 11.3 | 7.9 | 15.0 | 15.0 | 15.0 | 30.0 | 20.0 | 21.0 |

Note:
RSr: Rural seniors; RYa: Rural Yadults: USr: Urban seniors
$D_{0.5}$ (mile): 50th percentile of travel distance; $D_{0.75}$ (mile): 75th percentile of travel distance
$T_{0.5}$ (min): 50th percentile of travel time; $T_{0.75}$ (min): 75th percentile of travel time

Within a specific travel distance, $F_{D,RSr}(x) < F_{D,USr}(x)$ for all the travel purposes of study. That is, rural seniors can access fewer intended destinations than their urban counterparts within the same travel distance. For example, within 15 miles rural seniors reach 52.8% of their destinations for medical services, whereas this percentage for urban seniors is 82.9%. Similarly, given a specified travel time limit, $F_{T,RSr}(x) < F_{T,USr}(x)$ for the purposes of going home, accessing medical services, shopping, and transporting someone. That is, rural seniors experience drawbacks while accessing intended destinations than their counterparts in urban areas within the same travel time. For example, rural seniors reach 69.0% of their shopping destinations within 20 minutes, whereas urban seniors can reach 76.1%. Differences in their travel distance and travel time distributions are verified by the Kolmogorov-Smirnov (KS) test (p-value = 0).

Hypothesis H1 also states that aging is a factor that alters the intended destinations of rural seniors, which in turn changes their accessibility. To determine the direct impact of aging on accessibility, rural seniors are compared to rural Yadults with respect to their travel distance and travel time. For work commute, home returning, and recreational activities, $F_{D,RSr}(x) > F_{D,RYa}(x)$ at any given travel distance, and $F_{T,RSr}(x) > F_{T,RYa}(x)$ at any given travel duration. Those distinctions are verified by the KS test (p-value=0). For example, the observations pertaining to the purpose of work commute indicate that rural seniors intend to take job opportunities that are spatially and temporally closer to their homes than Yadults in rural areas.

Rural seniors are supposed to travel for a longer distance to access medical services, as compared to rural Yadults. For example, the 75th percentile travel distance to medical services is 22.9 miles for rural seniors and 20.4 miles for rural Yadults, as shown in Table 1. The difference in their travel distance CDFs is further verified by the KS test (p-value = 0.22), suggesting that these two groups have a moderate level of difference in travel distance to medical services. We conjecture that rural seniors may require special medical services more likely at farther destinations (e.g., in urban areas with well-developed medical services), but not for Yadults. Furthermore, in contrast with Yadults, rural seniors also have to travel for a longer time to access intended medical services. For example, Table 1 shows that 75% of rural seniors' trips to medical services are within 45.0 minutes, but it is 32.0 minutes for rural Yadults. The results indicate a restraint on accessing medical services for seniors than Yadults in rural areas both from the travel time and travel distance aspects.

For transporting someone, the disparity in travel distance CDFs between seniors and Yadults in rural areas is statistically significant (p-value of KS test is 0). Table 1 further shows the 75th percentile travel distance for rural seniors is 14.9 miles, whereas it is 11.3 miles for rural Yadults. Meanwhile, the difference in travel time between these two groups is also evident. The longer travel distance and travel time for rural seniors to transport someone to intended destinations is probably associated with the fact that rural seniors are more available than rural Yadults in providing transportation to others whose intended destinations are farther from their homes.

In summary, rural seniors encounter restricted accessibility for accessing medical services and assisting others. Nevertheless, it is noteworthy that they do not face equivalent limitations in activities such as returning home, work commute, shopping, or recreational pursuits. This distinction can be attributed to the special needs and willingness of this group to access some services or resources.

**The Impact of Aging and Rural Residency to Mobility.** The hypothesis H2 asserts that the reduced mobility among rural seniors stems from more restricted choices of their preferred transportation modes. In verifying this hypothesis, the study first analyzes the distribution of trips by transportation modes and traveler groups, as shown in Table 2. The marginal distribution of trips by transportation modes shows that automobiles are the most preferred mode, fulfilling 89.01% of trips. The frequency distribution of trips by transportation modes varies among the three groups according to the chi-squared contingency test(p-value=0). Rural seniors heavily rely on automobiles, which are used for 93.9% of their trips (=10.45%/11.13%). Walking is the secondary transportation mode for rural seniors, which fulfills 5.1% (=0.57%/11.13%) of their trips. Other modes count for an almost negligible amount. Although automobiles are still the primary transportation mode for urban seniors, the percentage of trips via automobiles is 84.3% (=38.32%/45.48%), 10.2% less compared to rural seniors. Besides automobiles, walking and transit are also their choices, which count for 11.3% (5.14%/45.48%) and 3.4% (=1.53%/45.48%) of their trips, respectively. What's more, the frequency of the trips among the transportation modes is similar between rural Yadults and rural seniors except that the former has a slightly lower proportion (92.8%) of trips using automobiles and a sensibly higher proportion (0.7%) of trips by transit. The heavy reliance on automobiles exposes rural seniors to the risk of compromised mobility when driving becomes unsuitable for them and fewer people can provide transportation to them.

Table 2. Distribution of trips by transportation modes and groups.

|  | Automobile | Walking | Transit | Bicycle | Others | Total |
|---|---|---|---|---|---|---|
| Rural Seniors | 10.45% | 0.57% | 0.04% | 0.01% | 0.06% | 11.13% |
| Rural Yadults | 40.25% | 2.24% | 0.32% | 0.14% | 0.45% | 43.39% |
| Urban Seniors | 38.32% | 5.14% | 1.53% | 0.24% | 0.25% | 45.48% |
| Total | 89.01% | 7.95% | 1.90% | 0.39% | 0.75% | 100.00% |

The study further calculates the measures of mobility, including travel frequency, travel speed, and time to public transit, for the three groups of travelers, which are summarized in Table 3. From the table, one can find that rural seniors have an average of 2.98 trips per person per day, the lowest among the three groups. The low travel frequency of rural seniors indicates a lower level of willingness they possess for travel. The travel speed of rural seniors is higher than urban seniors, which is attributed to the fact rural seniors are usually farther from their intended destinations than their counterparts in urban areas (Pucher and Renne 2005), which necessitates high-speed travel. However, the travel speed of rural seniors is lower than rural Yadults, evident that aging negatively influences rural travelers' mobility. The faster travel speed of in rural areas is faster than people in urban areas should not lead to the conclusion that mobility is higher in rural areas. Recognizing the constrained transportation choices and the restricted availability of public transit services for residents in rural areas is essential. This circumstance consequently amplifies the predominant reliance on automobiles in rural areas. Surprisingly, rural seniors require less time to reach transit stations, with an average of 7.29 minutes, in contrast to the other groups. It may seem contradictory to our initial hypothesis. However, rural seniors demonstrate a reduced preference for public transit (see Table 2). This shorter average time to the public transit pertains to only a small portion of trips.

Table 3. Measures of travel mobility.

|  | Frequency (per person per day) | Travel speed (mph) | | | Avg. time to public transit (min) |
|---|---|---|---|---|---|
|  |  | Automobile | Public transit | Non-motor |  |
| Rural Seniors | 2.98 | 27.6 | 19.2 | 5.7 | 7.29 |
| Urban Seniors | 3.24 | 20.0 | 11.6 | 3.8 | 7.85 |
| Rural Yadults | 3.31 | 30.8 | 21.7 | 6.2 | 10.89 |

In summary, aging and rural residency have been factors contributing to the reduced mobility among rural seniors. Aging is the main reason for the decreased travel frequency, and the lower density of opportunities, resources, and services in rural areas leads to their reliance on automobiles. Although automobiles meet their need for fast-speed travel, the heavy reliance on this mode without alternatives will cause a mobility crisis for this group if this preferred travel mode becomes infeasible.

**The Impact of Mobility on Accessibility.** Hypothesis 3 assumes that a higher level of travel mobility effectively increases accessibility, which is well discussed in the literature. This study attempts to verify this relationship using the NHTS dataset. Table 4 presents the 75th percentiles of travel distances, travel times, and travel speeds for the four travel modes associated with different travel purposes. Statistics in the table indicate long-distance trips rely on automobiles and public transit and, walking and riding bicycles are chosen for trips within short distances. The variation of trip distance CDF across those travel modes is further verified by the KS test (p-value=0).

**Table 4. Statistics of travel distance, travel time, and travel speed by transportation modes and travel purposes**

| Purpose | $D_{0.75}$ (mile) | | | | $T_{0.75}$ (min) | | | | $S_{0.75}$ (mph) | | | |
|---|---|---|---|---|---|---|---|---|---|---|---|---|
| | Walk | Bicycle | Auto | Transit | Walk | Bicycle | Auto | Transit | Walk | Bicycle | Auto | Transit |
| Home | 1.0 | 2.6 | 10.6 | 10.6 | 26.0 | 30.0 | 25.0 | 60.0 | 3.3 | 7.8 | 31.3 | 14.0 |
| Work | 0.6 | 3.2 | 15.6 | 14.0 | 15.0 | 30.0 | 30.0 | 60.0 | 3.4 | 8.7 | 35.9 | 15.9 |
| Medical | 0.6 | 2.2 | 13.5 | 7.8 | 15.0 | 25.0 | 30.0 | 60.0 | 2.8 | 6.5 | 31.9 | 12.6 |
| Shopping | 0.5 | 1.5 | 6.7 | 7.2 | 15.0 | 25.0 | 18.0 | 50.0 | 3.2 | 6.0 | 27.7 | 12.4 |
| Recreational | 0.8 | 2.5 | 11.5 | 9.9 | 19.0 | 30.0 | 27.0 | 46.0 | 2.0 | 9.4 | 32.8 | 15.5 |
| Transport someone | 0.6 | 1.2 | 8.4 | 9.0 | 15.0 | 15.0 | 20.0 | 45.0 | 3.2 | 7.5 | 30.4 | 19.7 |

Note:
$D_{0.75}$: 75th percentile of travel distance; $T_{0.75}$: 75th percentile of travel time; $S_{0.75}$: 75th percentile of travel speed

While automobiles and transit are both options for long-distance trips, the former offers a higher level of mobility than the latter. The 75th percentile speed of automobiles is at least twice of the transit for all purposes except for transporting someone, and the 75th percentile travel time of automobiles is 41%~64% less than the transit. Automobiles move travelers to farther destinations of work, medical services, and recreational facilities than the public transit does, which is probably attributed to the higher level of mobility with automobiles. Notably, the 75th percentile travel distance to medical services via automobiles is 13.5, but it is 7.8 for the transit. Yet, for the purposes of shopping and transporting someone, the transit moves people to their slightly farther (0.5~0.6mi) destinations than automobiles. Table 2 shows that trips via automobiles are 89.01% whereas those via public transit are 1.9%. The distinctly different proportions of trips by those two transportation modes are probably a result of the public transit's lower level of mobility than automobiles.

While walking and riding bicycles are both short-distance transportation modes, the latter offers a higher level of mobility than the former. The 75th percentile speed of riding bicycles is 88%~370% faster than walking. Consequently, bicycles move travelers to destinations 100%~433% farther than walking. However, Table 2 shows only 0.39% of trips use bicycles, significantly lower than walking (7.95%).

To sum up, automobiles offer the highest level of mobility among the four transportation modes, making it the dominating mode of transportation in the United

States for all travel purposes. Transit, as an alternative to automobiles for long-term travel, offers a lower level of mobility and thus transports travelers to closer destinations for certain travel purposes like accessing medical services. Riding bicycles provides a higher level of mobility than walking for short-distance travels, bringing travelers to farther destinations at a faster speed than walking can reach. However, the proportion of trips by riding bicycles is significantly lower than walking, indicating certain constraints such as biking infrastructure prevent travelers from switching from walking to riding bicycles. The observations underscore the fact that the higher level of service road system and the mass rapid transit system can improve mobility and, in turn, accessibility.

**The Impact of Aging and Rural Residency to Accessibility via Mobility.** The hypotheses H2 and H3 together indicate that mobility is a mediator on a causal pathway illustrated in Figure 1. That is, aging and rural residency of travelers raise mobility issues, which in turn limit rural seniors' accessibility to certain desired opportunities, resources, and services. This study further verified the mediator role of mobility by examining the travel distance and travel time of the three traveler groups under selected combinations of travel purposes and transportation modes. Table 5 summarizes the 75th percentiles of travel distance and travel time.

**Table 5. Statistics of travel distance and travel time, by combinations of transportation modes and travel purposes among rural seniors (RSr), rural Yadults (RYa), and urban seniors (USr).**

|     | Shopping | | | | | | Medical Service | | | | | | Transport Someone | | | |
|-----|---|---|---|---|---|---|---|---|---|---|---|---|---|---|---|---|
|     | Walk | | Auto | | Transit | | Walk | | Auto | | Transit | | Auto | | Transit | |
|     | $D_{0.75}$ | $T_{0.75}$ | $D_{0.75}$ | $T_{0.75}$ | $D_{0.75}$ | $T_{0.75}$ | $D_{0.75}$ | $T_{0.75}$ | $D_{0.75}$ | $T_{0.75}$ | $D_{0.75}$ | $T_{0.75}$ | $D_{0.75}$ | $T_{0.75}$ | $D_{0.75}$ | $T_{0.75}$ |
| RSr | 0.34 | 10.0 | 10.8 | 20.0 | 17.2 | 40.0 | 0.4 | 15.0 | 22.2 | 45.0 | 22.2 | 55.0 | 14.9 | 30.0 | 19.0 | 50.0 |
| RYa | 0.58 | 20.0 | 10.5 | 20.0 | 14.0 | 27.0 | 5.1 | 25.0 | 20.6 | 34.0 | 15.0 | 26.0 | 11.6 | 20.0 | 4.3 | 15.0 |
| USr | 0.44 | 15.0 | 5.2 | 16.0 | 4.7 | 48.0 | 0.7 | 20.0 | 11.9 | 30.0 | 10.3 | 50.0 | 8.1 | 22.0 | 16.5 | 45.0 |

Note:
$D_{0.75}$:75th percentile of travel distance (mile) ; $T_{0.75}$:75th percentile of travel time (min)

Walking is the major transportation mode for the short-distance movement of rural seniors. 75% of rural seniors' trips to shopping centers by walking are within 0.34 miles or 10 minutes. These statistics are 0.58 miles and 20 minutes for rural Yadults, and 0.44 miles and 15 minutes for urban seniors. This phenomenon is due to rural seniors' preference for shorter travel distances and time if shopping centers are nearby. However, automobiles, as the dominating transportation mode for long-distance movement, have limited accessibility for senior residents in rural areas to shop, due to the higher travel cost than urban seniors. Meanwhile, 75% of rural seniors' trips to shopping centers via automobiles are within 10.8 miles and 20 minutes. The statistics are 10.5 miles and 20 minutes for rural Yadults. Therefore, the data evidence indicates that they confront equitable accessibility to shopping centers via automobiles.

This study has found that rural seniors need to access a larger percentage of medical services that are at farther distances and require a longer time to reach than rural Yadults and urban seniors. Table 5 shows that their 75th percentile travel distance to medical services using automobiles is 22.2 miles and the 75th percentile travel time is 45 minutes. These statistics are 20.6 miles and 34 minutes for rural Yadults and 11.9 miles and 30 minutes for urban seniors. The comparison indicates that automobiles, as a dominating transportation mode, offer a lower level of mobility for rural seniors in accessing medical services than for rural Yadults and urban seniors. Rural seniors also undertake longer travel distance ($D_{0.75} = 22.2$) and time ($T_{0.75} = 55.0$) to access medical services if taking transit, due to its limited mobility among rural seniors. Simultaneously, walking is also a transportation mode with a reduced level of mobility for rural seniors than their counterparts. The 75th percentile of travel distance is 0.4 miles and the 75th percentile of travel time is 15 minutes for the former, whereas those statistics are 5.1 miles and 25 minutes for rural Yadults; and 0.7 miles and 20 minutes for urban seniors. Improving the mobility level of automobiles and transit for rural seniors would provide them with more equitable accessibility to medical services as compared to other demographic groups.

The final focus of understanding the mediator role of mobility centers on the purpose of transporting others. As found from the study of H1, rural seniors have a higher percentage of trips that spend longer time and travel for a longer distance to reach the destinations for assisting someone than rural Yadults. This accessibility difference between them is particularly distinct in using the transit. Table 5 shows the 75th percentile of travel distance is 19.0 miles and the 75th percentile of travel time is 50.0 minutes for rural seniors, and those statistics are 4.3 miles and 15 minutes for rural Yadults. Rural Yadults demonstrate different behavior than seniors in choosing transportation modes for transporting others, indicating the mobility levels of automobiles and transit are different among these groups. Meanwhile, the significant distinction between rural seniors and urban seniors is from the utilization of automobiles. As shown in Table 5, the 75th percentile of travel distance is 14.9 miles, and the 75th percentile of travel time is 30.0 minutes for rural seniors, but 8.1 miles and 22.0 minutes for urban seniors. This disparity of accessibility is attributed to fewer travel mode alternatives to automobiles. Specifically, poor public transit facilities and low-density distribution of opportunities, resources, and services in rural areas lead to the heavy reliance on automobiles.

In summary, mobility is a mediator through which aging can lower rural seniors' accessibility. Automobiles, as the dominating transportation mode for long-distance travel, offer a lower level of mobility to rural seniors than rural Yadults, especially in accessing medical services and transporting others. Walking is the major mode for short-distance movement, but it prevents rural seniors from accessing destinations that are a little farther and accessible by rural Yadults or urban seniors. Rural seniors have limited choices of transportation modes, for both long-distance and short-distance travel, making it more difficult to reach intended farther destinations.

**CONCLUSION**

This study presents a causal model delineating both the direct and indirect effects of aging and rural residency on travelers' accessibility to opportunities, resources, and services. In this model, mobility serves as a mediator through which the demographic attributes of rural seniors indirectly influence their accessibility. Descriptive statistics of the trip data of 2017 National Household Travel Survey support our hypotheses, confirming the presence of the proposed causal relationships.

An immediate step following this study is to estimate the coefficients that quantify the strengths and directions of the causal relationships. Given such a model, the effectiveness of improving equitable accessibility for rural seniors by enhancing their mobility can be estimated. Additionally, the causal model can be further improved by integrating additional causal relationships. Beyond aging and rural residency, various additional factors, such as land use, traffic congestion, opportunity density, and infrastructure density, impact travel mobility and accessibility. It is worth noting that using distance to evaluate accessibility has limitations due to the effect of distance decay, which represents the level of reluctance to travel long distances among regions. Future studies could explore modified indicators for accessibility and refine the construction of a comprehensive causal model to address these considerations.

**ACKNOWLEDGEMENT**

This research was partially funded by the Rural, Equitable and Accessible Transportation (REAT) Center, a Tier-1 University Transportation Center (UTC) funded by the United States Department of Transportation (USDOT), through agreement number 69A3552348321. The contents of this paper reflect the views of the authors. USDOT assumes no liability for the contents or use thereof.